# High Pressure Effects on Thermal Properties of MgO

Iris Inbar and R. E. Cohen

Geophysical Laboratory and Center for High Pressure Research, Carnegie Institution of Washington, Washington DC 20015

**Abstract.**
Using the non-empirical Variational Induced Breathing (VIB) model, the thermal properties of periclase (MgO) under high pressures and temperatures are investigated using molecular dynamics, which includes all anharmonic effects. Equations of state for temperatures up to 3000K and pressures up to 310 GPa were calculated. Bulk modulus, thermal expansivity, Anderson-Grüneisen parameter, thermal pressure, Grüneisen parameter and their pressure and temperature dependencies are studied in order to better understand high pressure effects on thermal properties. The results agree very well with experiments and show that the thermal expansivity decreases with pressure up to about 100 GPa ($\eta=0.73$), and is almost pressure and temperature independent above this compression. It is also effected by anharmonicity at zero pressure and temperatures above 2500K. The thermal pressure changes very little with increasing pressures and temperatures, and the Grüneisen parameter is temperature independent and decreases slightly with pressure.

## Introduction

The density of lower mantle minerals at high pressures and temperatures is crucial for interpreting seismological data in terms of mineral constituents. For the first time all anharmonic effects up to extreme pressures and temperatures are being addressed using a non-empirical, first principles ionic model. We study MgO here since 1) We have an extremely well characterized and accurate potential for MgO; 2) it is an end member of magnesiowüstite (Mg,Fe)O thought to be a major constituent of the lower mantle; 3) MgO has no phase transitions or soft modes in the region of interest and therefore should show generic oxide character; and 4) MgO is the simplest oxide; other dense minerals whose compression is homogeneous should behave similarly.

Previous studies using non-empirical potentials [Isaak et al. 1990; Wolf and Bukowinski, 1988], empirical potentials [Agnon and Bukowinski, 1990, Reynard and Price, 1990] and thermodynamic models [Anderson et al. 1992, 1993] all concluded that the thermal expansivity becomes temperature independent at high pressures. All studies are consistent with an Anderson-Grüniesen parameter $\delta_T = (\frac{\partial ln\alpha}{\partial lnV})_T$ equal to about 5.0 at zero pressure which decreases with pressure. How anharmonicity might effect these results and how $\delta_T$ behaves with pressure and temperature, as well as other thermal quantities like $K_T$, $\gamma$ and $P_{TH}$ is addressed here.

## Method

We have developed a massively parallel molecular dynamics code for the connection machine CM-5, using the VIB model [Wolf and Bukowinski, 1988] and periodic boundary conditions. We have used supercells of 64 and 216 atoms, and performed constant E, N and V simulations; no scaling was used after equilibration. Results of P and T were obtained by averaging over 4-10 ps after the system has equilibrated, which took 1-2 ps. The runs were sufficiently long so that statistical errors were smaller than the size of the symbol. The Variational Induced Breathing (VIB) potential is based on density functional theory which states that all the ground state properties are determined by the charge density alone, which is modeled here by overlapping ionic charge densities. Since oxygen ions are unstable in the free state, they are surrounded by a Watson sphere of charge +2. The Watson spheres are allowed to relax spherically at each time step to minimize the total energy. In the related PIB model the Watson sphere radius is chosen to give the Madelung potential at the ion site. The Local Density Approximation (LDA) is used to calculate the total energy which is written as the sum of the self-energy of each ion, the electrostatic energies, the overlap kinetic energy (calculated using the local Thomas-Fermi electron gas functional) and exchange and correlation energies, which are all functionals of the charge density. Further discussion of the method can be found in Cohen and Gong [1994], who studied the melting of MgO using PIB rather than VIB, and a finite cluster rather than periodic boundary conditions.

## Results and Discussion

We find excellent agreement between our equation of state parameters and experimental values (Table 1). The temperature variation of the thermal expansivity is shown in Fig. 1a at pressures up to 310 GPa. Since we use classical MD, and quantum thermal and zero point corrections are neglected, we expect the results to be accurate at temperatures above the Debye temperature ($\Theta_D$=945K for MgO). Above the Debye temperature the thermal expansivity agrees well with experimental data. Previous non-empirical studies have never included all effects of anharmonicity; here at high temperatures and zero pressure we find a nearly linear thermal expansivity with temperature compared with the diverging form found by Isaak *et al.* [1990] which employed the quasi-harmonic approximation. At higher pressures (100GPa) and high temperatures we do not find significant differences between our results and the quasi-harmonic results implying therefore that at high pressures anharmonicity does not play a major role. At high pressures (Fig. 1b) we also find excellent agreement between the results of this study and experimental data. Duffy and Ahrens [1993] measured the thermal expansivity at $\overline{T} = 1800$K and pressures ranging between 160 and 200 GPa to be a constant within experimental errors. These results agree extremely well with our values for the T=1880 K isotherm at these pressures. We further predict the thermal expansivity to be pressure and also temperature independent at higher pressures and at higher temperatures as previous non-empirical studies have not studied the effects of such extreme pressures, up to 310 GPa, on $\alpha$. The thermal expansivity decreases with pressure up to about 100-150GPa ($\eta$=0.73-0.65) at which point it becomes pressure and temperature independent.

Shown in Fig. 2 as a function of pressure is the Anderson-Grüneisen parameter, $\delta_T = (\partial ln\alpha/\partial ln V)_T$. For $\delta_T$ at ambient conditions, we obtain $\delta_T$=4.8, in excellent agreement with experiment [Chopelas and Boehler, 1989]. With increasing pressure the value of the Anderson-Grüneisen parameter decreases and becomes temperature independent, much like Agnon and Bukowinski's [1990] empirical model. Anderson and Isaak [1993] have suggested that $\delta_T = \delta_{T_0}\eta^\kappa$, with $\kappa$ =1.4 and independent of T and $\eta$. We find that this is correct only up to about $\eta$=0.8 (50 GPa), at which point this begins to diverge from our results (Fig. 2).

The quantity $\alpha K_T$ is important in determining other thermodynamic properties. The bulk modulus $K_T$ is also almost temperature independent at high pressures (Fig 3) and is consistent with the above results as $(\frac{\partial \alpha}{\partial P})_T \equiv \frac{1}{K_T^2}(\frac{\partial K_T}{\partial T})_P$. At zero pressure we get very good agreement with experiment and at high temperatures there is no difference between our results and the quasi-harmonic results implying anharmonicity does not effect $K_T$.

The Grüneisen parameter $\gamma$ depends on $\alpha K_T$ via $\gamma = \frac{\alpha K_T V}{C_V}$, and its temperature and pressure dependence are shown in Fig 4a and b, respectively. We find that the Grüneisen parameter is temperature independent in accord with experimental results and decreases slightly as a function of pressure, which is similar to other theoretical models [Agnon and Bukowinski, 1990] and calculations [Isaak et al., 1990b]. Here too, the effects of not including anharmonicity are shown by the diverging form of the dashed line corresponding to the results of Isaak et al. who used the quasi-harmonic approximation.

The thermal pressure along an isochore is defined as $P_{TH} = \int_o^T (\partial P/\partial T)_V \, dt = Const. + \alpha K_T T$. We calculated the pressure dependence of the thermal pressure at different temperatures above the Debye temperature. As pressure increases, the thermal pressure increases, up to about 100 GPa ($\eta$=0.73), at which point the thermal pressure becomes pressure independent. In order to understand this phenomenon we can look at $(\frac{\partial (\alpha K_T)}{\partial P})_T = \alpha(K_T^{'} - \delta_T)$. Our results show that while $K_T^{'}$ is slightly smaller than $\delta_T$ at zero pressure ($K_T^{'}$ = 4.68, $\delta_T$=4.8 and $K_T^{'}$=$\delta_T$ at 20 Gpa), $\delta_T$ decreases much faster than $K_T^{'}$ at pressures up to 150 GPa ($\delta_T$ decreases exponentially and $K_T^{'}$ decreases linearly). At this point $\delta_T$ becomes almost pressure independent while $K_T^{'}$ continuously decreases until at 250 GPa $K^{'}=\delta_T$.

## Conclusions

These *ab initio* calculations yield values for the thermal expansivity which are in very good agreement with available experimental data and show that with increasing pressure the thermal expansivity decreases, up to about 120 GPa ($\eta$=0.7) at which pressure it becomes pressure and temperature independent. Anharmonicity effects are shown to be important only at zero pressure and temperatures above 2500K. The Anderson-Grüneisen parameter $\delta_T$ follows the same behavior; it decreases with increasing pressure and become temperature independent at high pressures. The bulk modulus is found to be temperature independent at high pressures and unaffected by anharmonicity. The Grüneisen parameter $\gamma$ and the thermal pressure are also calculated at various pressures and temperatures and are also found to be almost temperature and pressure independent, which is consistent with the assumption made by Birch [1952] that $\alpha K_T$ is independent of density in the deep earth. Previous non-empirical studies have never found such good agreement with experiment.

**Acknowledgments.** We would like to thank Joe Feldman, T. Duffy, D. G. Isaak, R. Hazen and R. T. Downs for many helpful and stimulating discussions.# References

A. Agnon and M. S. T. Bukowinski, Thermodynamic and elastic properties of a many-body model for simple oxides, *Phys. Rev. B* **41**, 7755 (1990)

O. L. Anderson, D. Isaak and H. Oda, High-temperature elastic constant data on minerals relevant to geophysics, *Reviews of Geophysics* **30**, 57 (1992)

O. L. Anderson, H. Oda, A. Chopelas and D. G. Isaak, A thermodynamic theory of the Grüneisen parameter ratio at extreme conditions: MgO as an example, *Phys. Chem. Minerals* **19**, 369 (1993)

O. L. Anderson and D. G. Isaak, The dependence of the Anderson-Grüneisen parameter $\delta_T$ upon compression at extreme conditions, *J. Phys. Chem. Solids* **54**, 221 (1993)

F. Birch, Elasticity and constitution of the earth's interior, *J. Geophys. Res.* **57**, 227 (1952)

Z. P. Chang and G. R. Barsch, Pressure dependence of the elastic constants of single crystal magnesium oxide *J. Geophys. Res.* **74**, 3291 (1969)

A. Chopelas and R. Boehler, Thermal expansion measurements at very high pressure, systematics, and a case for a chemically homogeneous mantle *Geophys. Res. Lett.* **16**, 1347 (1989)

A. Chopelas and R. Boehler, Thermal expansivity in the lower mantle, *Geophys. Res. Lett.* **19**, 1983 (1992)

R. E. Cohen and Z. Gong, Melting and melt structure of MgO at high pressures, *Phys. Rev. B* **50**, 12301 (1994)

T. Duffy, R. J. Hemley and H. Mao, Equation-of-state and shear strength of magnesium oxide to 227 GPa, *Phys. Rev. Lett.* **74**, 1371 (1995)

T. S. Duffy and T. J. Ahrens, Thermal expansion of mantle and core materials at very high pressures, *Geophys. Res. Lett.* **20**, 1103 (1993)

D. G. Isaak, O. L. Anderson and T. Goto, Measured elastic moduli of single crystal MgO up to 1800 K, *Phys. Chem. Minerals* **16**, 704 (1989)

D. G. Isaak, R. E. Cohen and M. J. Mehl, Calculated elastic and thermal properties of MgO at high pressures and temperatures, *J. Geophys. Res.* **95**, 7055 (1990)

H. H. Mao and P. M. Bell, Equation-of-state of MgO and Fe under static pressure conditions, *J. Geophys. Res.* **84**, 4533 (1979)

B. Reynard and G. D. Price, Thermal expansion of mantle minerals at high pressures - a thermodynamic study, *Geophys. Phys. Lett.* **17**, 689 (1990)

Y. Sumino, O.L. Anderson and I. Suzuki, Temperature coefficients of elastic constants of single crystal MgO between 80 and 1300 K, *Phys. Chem. Minerals* **9**, 38 (1983)

D. Garvin, V. B. Parker and H. J. White, (Eds.) *CODATA Thermodynamic tables*, Hemisphere, Washington, D.C,
5

1987

Y. S. Touloukian, R. K. Kirby, R. E. Taylor and T. Y. R. Lee (Eds.) *Thermophysical Properties of Matter* vol.**13**, IFI/Plenum, New York, 1977

G. H. Wolf and M. S. T. Bukowinski, Variational stabilization of the ionic charge densities in the electron-gas theory of crystals: applications to MgO and CaO, *Phys. Chem. Minerals* **15**, 209 (1988)

Iris Inbar and R. E. Cohen, Carnegie Institution of Washington, Geophysical Laboratory and Center for High Pressure Research

5251 Broad Branch Rd, N.W. Washington, D.C. 20015

:

:

:

:

:

:

**Figure 1.** a). Temperature variation of the thermal expansivity at pressures up to 310 GPa. Lines for 0, 100, 200 and 310 GPa are results from this study, dashed line is from Isaak *et al.* [1990] +'s are data from Touloukian *et al.*, and *'s are data from Anderson *et al.* [1993a] at 100 GPa. b). Pressure variation of the thermal expansivity at different temperatures above the Debye temperature. Lines are results of this study, •'s are data from Duffy and Ahrens at 1800K, x's are from Chopelas and Boehler [1992] at 2000K.

**Figure 2.** Pressure Variation of $\delta_T$ at different temperatures. The o's are T=940K, +'s are T=1880K and *'s are T=2830K. $\delta_T$ decreases with pressure and becomes temperature independent at high pressures. x's represent Anderson-Isaak proposed fit of $\delta_T = \delta_{T_0} \eta^{1.4}$.

**Figure 3.** The temperature variation of the Bulk modulus for different pressures. The +'s are data from Anderson *et al.* [1992], and the solid line is from Isaak *et al.* [1990] using the PIB model and the quasi-harmonic approximation.

**Figure 4.** a). The temperature dependence of the Grüneisen parameter at zero pressure. The solid line represents results of this study, the dashed line is from calculations by Isaak *et al.* [1990] and the +'s are experimental results from Sumino, Isaak and Garvin. b). Pressure variation of the Grüneisen parameter at 1000K. Results from this study are shown by o's, +'s represent results from calculations by Isaak *et al.* [1990] and *'s are calculations from a model by Anderson *et al.* [1993a].

**Table 1.** Calculated Equation-of-state Parameters for MgO at zero pressure.

|  | T = 300 K | | | | | | This Study | | |
|---|---|---|---|---|---|---|---|---|---|
|  | Exp. | This Study | QH PIB[a] | QH VIB[b] | EM[c] | EM[d] | 940K | 1880K | 2830K |
| V [$A^3$] | 18.66[e] | 18.70 | 18.66 | 18.69 | 18.73 | - | 19.29 | 20.29 | 21.38 |
| $K_T$ [GPa] | 160.1[f] | 153.11 | 180.1 | 177 | 232 | 161.3 | 132.09 | 104.15 | 83.93 |
| $K_T'$ | 4.22[g] | 4.68 | 4.15 | - | - | - | 5.01 | 5.44 | 5.66 |
| $K_T''$ | - | -0.05 | -0.026 | - | - | - | -0.067 | -0.07 | -0.09 |
| $\alpha \times 10^{-6}$ [$K^{-1}$] | 32.0[h] | 38.79 | 23.9 | - | 19.4 | 31 | 44.56 | 52.41 | 58.75 |

a) QH:Quasi-harmonic, Isaak *et al.* [1990], b) Wolf and Bukowinski [1988], c) EM:Empirical model, Reynard and Price [1990], d) Agnon and Bukowinski [1990], e) Mao and Bell [1979], f) Sumino *et al.* [1983], g) Chang and Barsch [1969], h) Touloukian *et al.*[1977]

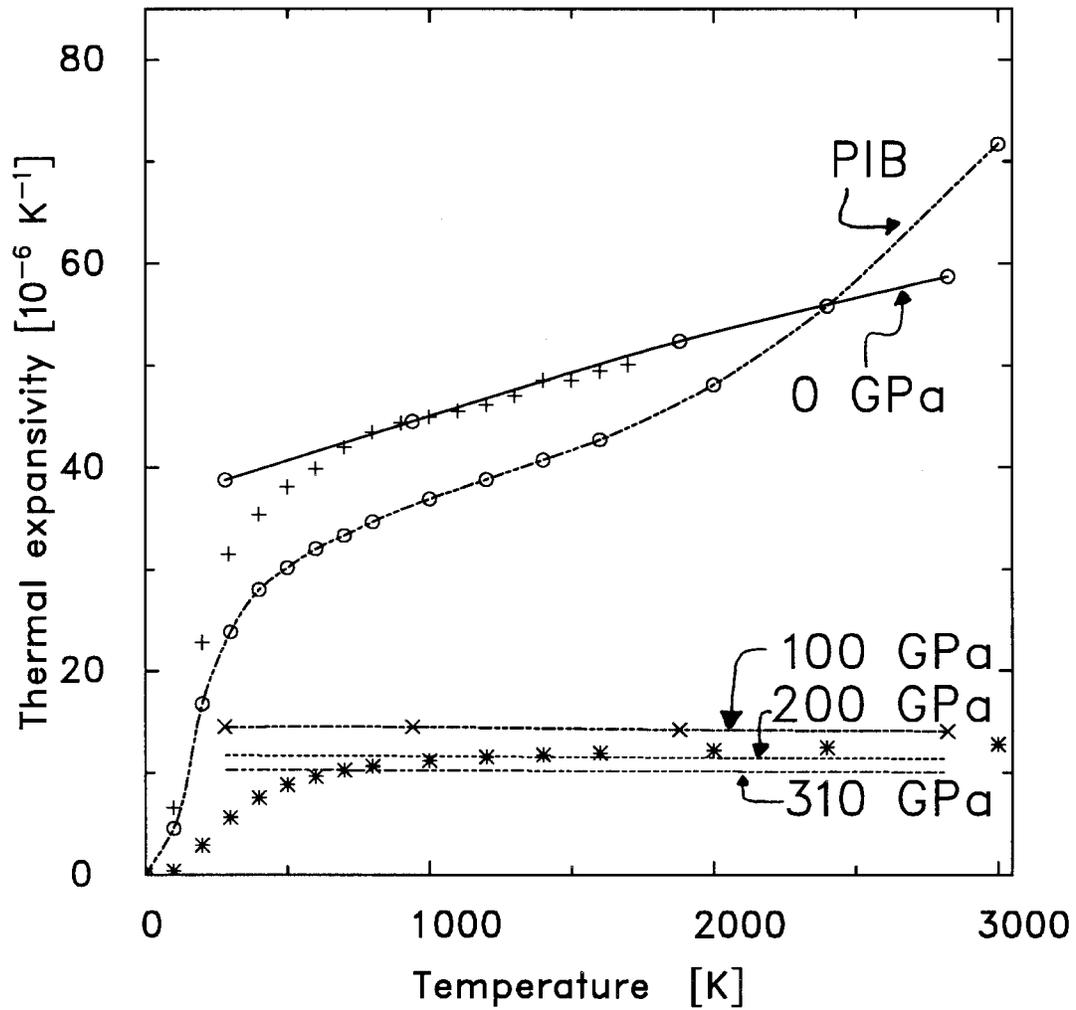

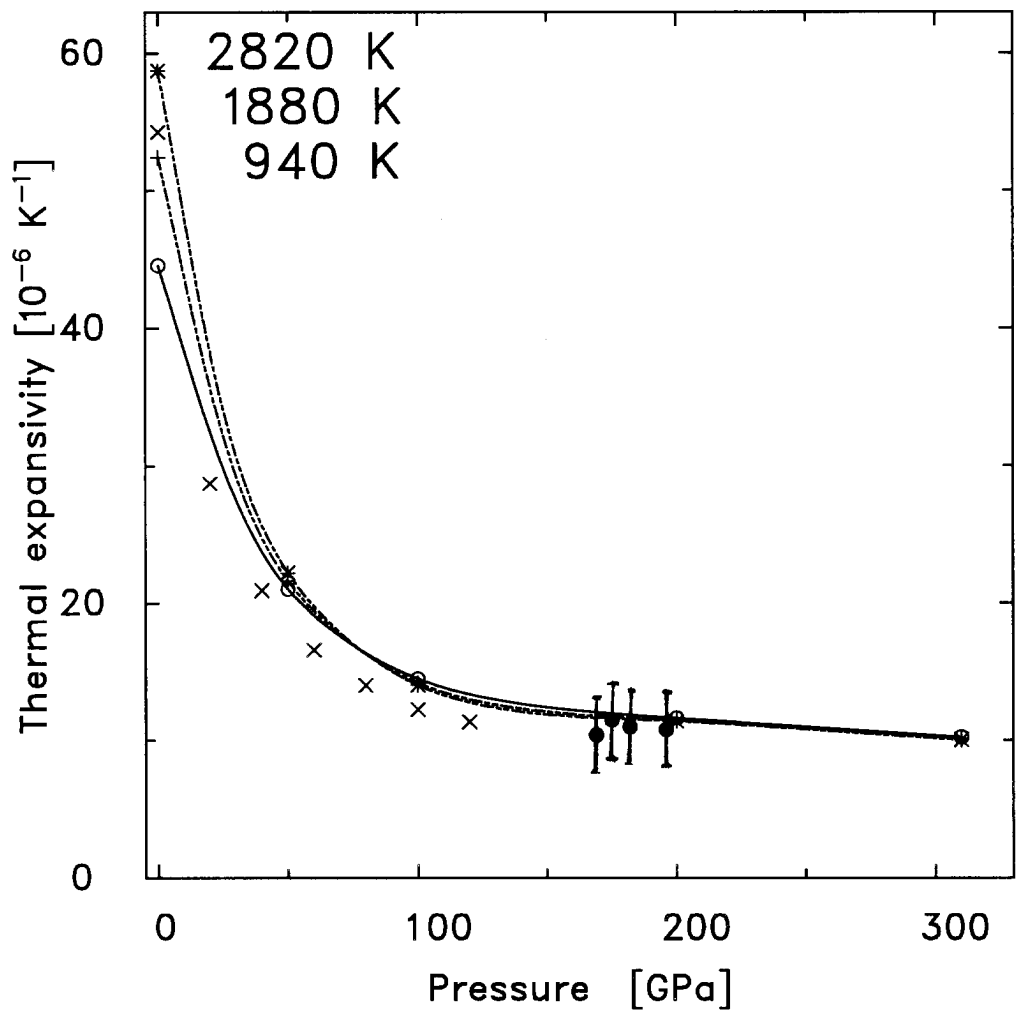

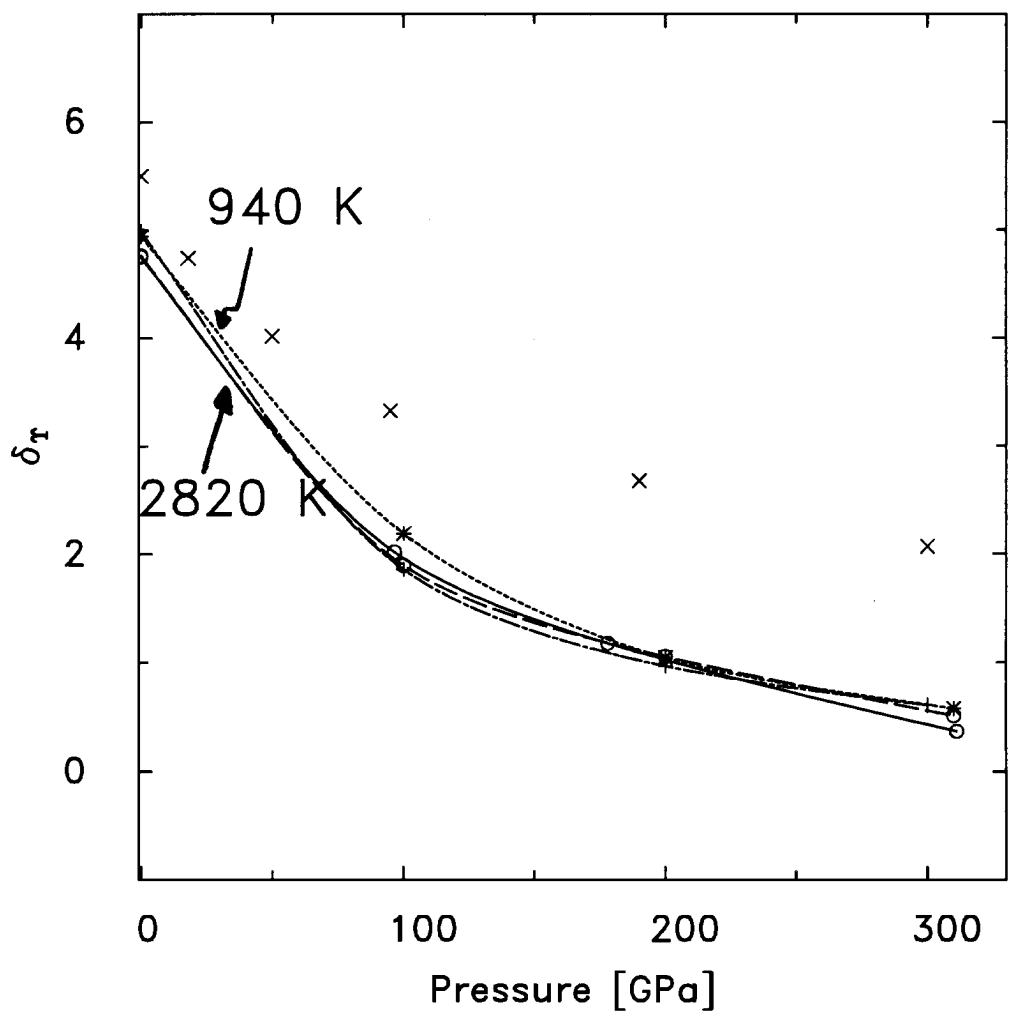

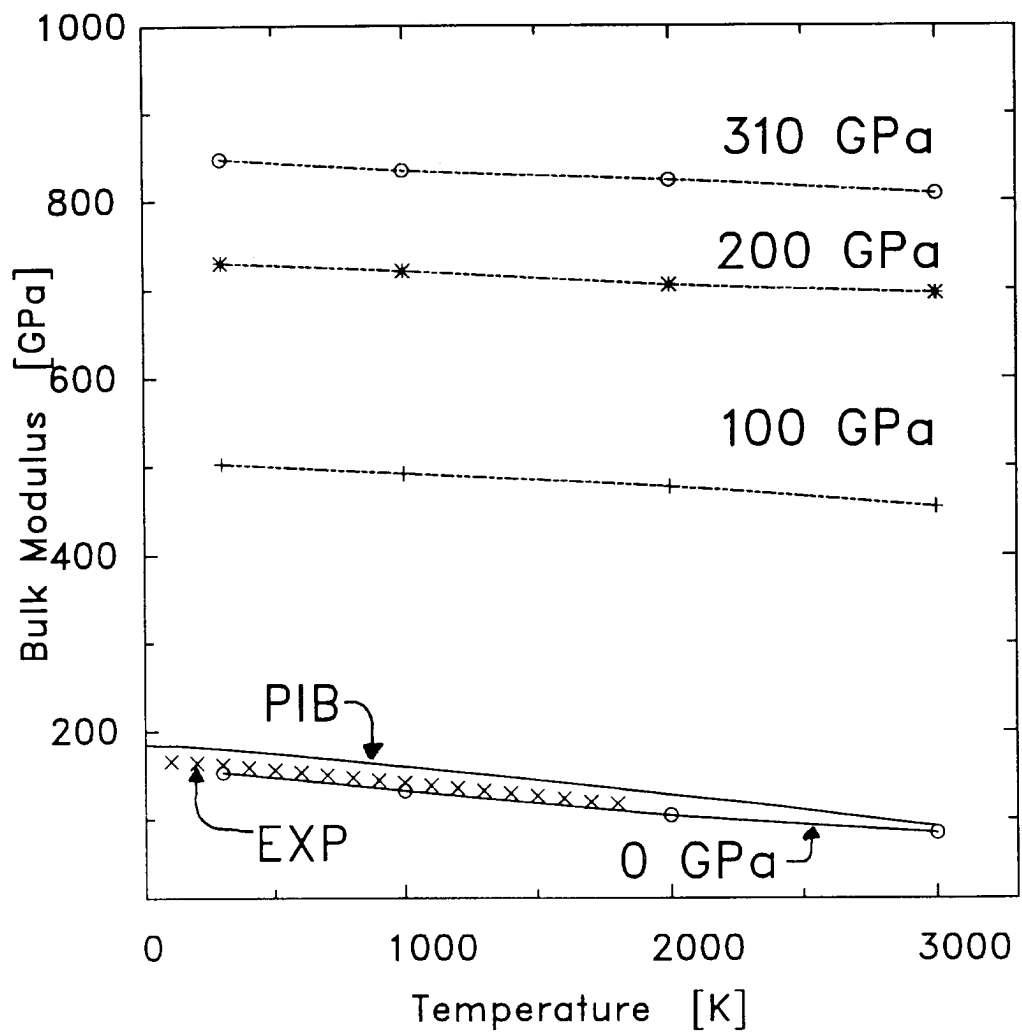

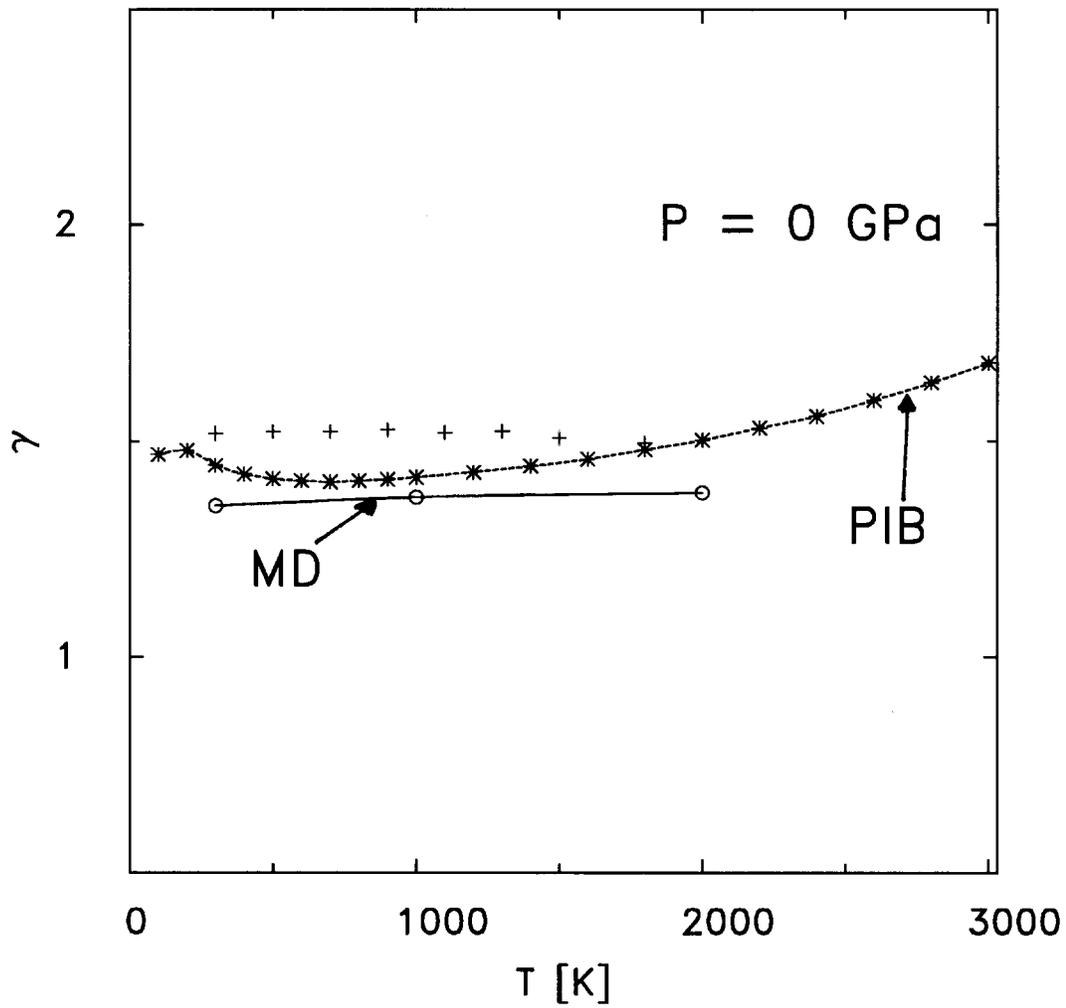

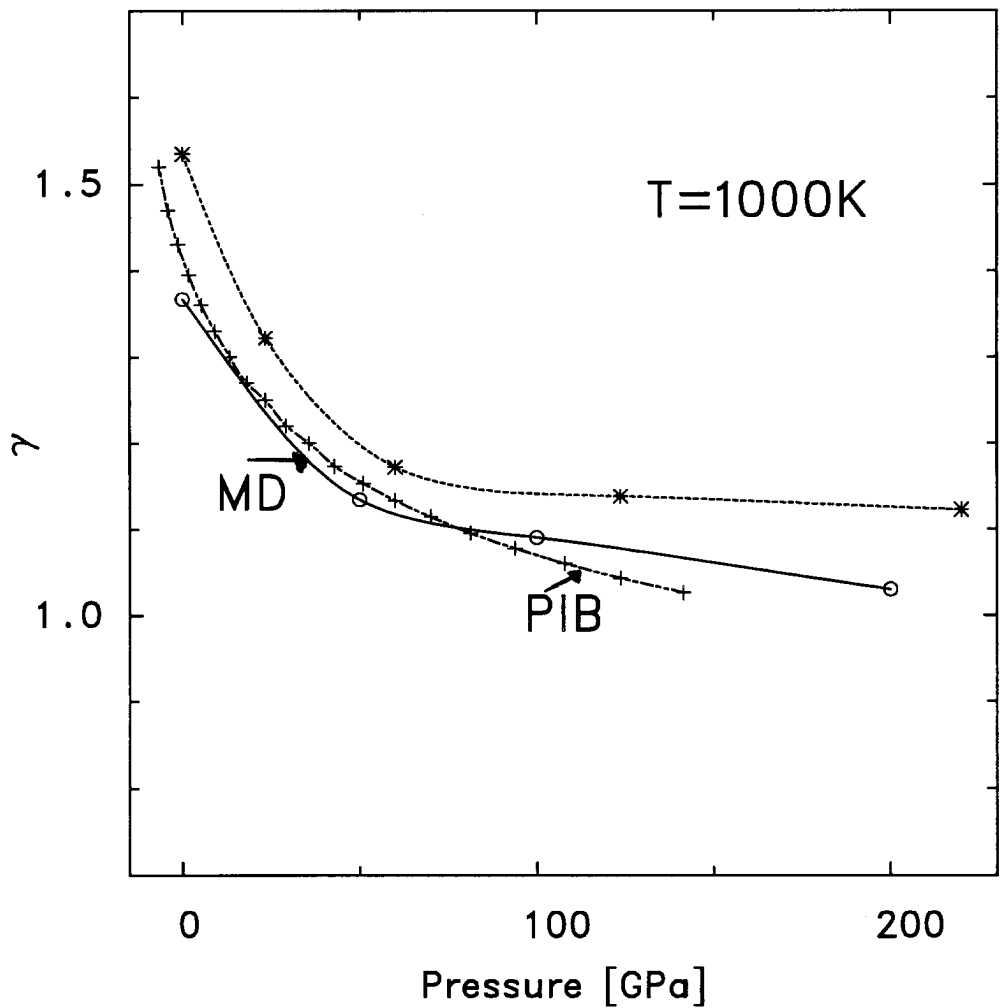